\documentclass[prl,twocolumn,amsmath,amssymb]{revtex4-2}

\usepackage{amsfonts,amsmath,amssymb,graphicx}
\usepackage[colorlinks=true,allcolors=blue]{hyperref}

\newcommand{\arXiv}[1]{\href{http://www.arXiv.org/abs/#1}{arXiv:#1}}
\renewcommand{\doi}[2]{\href{http://dx.doi.org/#2}{#1}}
\newcommand{\beq}{\begin{equation}}
\newcommand{\eeq}{\end{equation}}
\newcommand{\del}{\partial}
\newcommand{\sssty}{\scriptscriptstyle}

\newcommand{\de}{\delta}

\renewcommand{\Re}{\operatorname{Re}}
\renewcommand{\Im}{\operatorname{Im}}
\newcommand{\Tr}{\mathrm{Tr}}
\newcommand{\nn}{\nonumber}
\newcommand{\nodisplaybreaks}{\interdisplaylinepenalty=10000}

\begin{document}

\title{Analytic and numerical toolkit for the Anderson model in one dimension}

\author{Oleg Evnin\vspace{2mm}}
\affiliation{High Energy Physics Research Unit, Faculty of Science, Chulalongkorn University, Bangkok, Thailand \&\\
Theoretische Natuurkunde, Vrije Universiteit Brussel \& International Solvay Institutes, Brussels, Belgium}

\begin{abstract}
The Anderson model in one dimension is a quantum particle on a discrete chain of sites with nearest-neighbor hopping and random on-site potentials. 
It is a progenitor of many further models of disordered systems, and it has spurred numerous developments in various branches of physics.
The literature  does not readily provide, however, practical analytic tools for computing the density-of-states of this model when the distribution of the on-site potentials
is arbitrary. Here, supersymmetry-based techniques are employed to give an explicit linear integral equation whose solutions control the density-of-states.
The output of this analytic procedure is in perfect agreement with numerical sampling. By Thouless formula, these results immediately provide analytic
control over the localization length.

\end{abstract}

\maketitle

Anderson localization \cite{anderson} has been around for nearly 70 years, becoming one of the reference themes for studies of disorder.
It has found numerous applications to transport phenomena in solid state physics, as well as propagation of light and sound in disordered media \cite{phystoday}.

The simplest one-dimensional (1d) Anderson model can be visualized as a quantum particle that is allowed to hop between the nearest-neighbor sites of a chain,
while random potentials, with a prescribed probability distribution, are assigned independently to each site. Somewhat surprisingly, despite the amount
of attention this model has received, no simple analytic procedures are seen in the literature for computing its density-of-states (DoS) when the on-site disorder distributions are arbitrary given functions. The purpose of this Letter is to provide such a procedure. The derivation will start with the rather conventional framework of supersymmetry-based methods in application to disorder averaging \cite{Efetov, nucleus,FMsigma,Mirlin, textbooks}, but will depart from the standard expositions in the literature at an essential point, taking a shortcut to a simple, explicit result in the form of a linear integral equation. While the phenomenology of the Anderson model is understood in terms of the eigenvector localization over the lattice sites, rather than the DoS, the Thouless formula \cite{thouless} provides an expression for the localization length in terms of the DoS, and thus, the analytic solution presented here
will automatically control the key physical localization properties.

Integral equations have appeared in the literature for some related problems. First, there is a series of works starting with Dyson's classic paper on disordered oscillator chains \cite{Dyson}, see \cite{Schmidt,Halperin,Forrester} for some of the follow-up. The methodology and presentation of the final results in these works are rather different from what will be pursued here, and the specific way disorder is introduced likewise differs from the standard 1d Anderson model, but the final output is in the form of integral equations. Another important branch of literature starts with the classic works of Abou Chacra-Anderson-Thouless on Bethe lattices \cite{ACAT}, see \cite{nuclphysb,DerridaRodgers,BST} for some of the follow-up. These considerations are framed in relation to the `infinite-dimensional' physics of Bethe lattices, but some scrutiny of the formalism reveals that the derivations only use the treelike character of the graph and hence apply to the Bethe lattice with branching ratio 1, which is the 1d chain. The nonlinear integral equations relevant for the general case reduce for this case to linear equations, making contacts with the exposition here, which is easier to see in \cite{nuclphysb,DerridaRodgers}. Further comments on the history of the subject can be found in the supplemental material \cite{sppl}.

The construction presented below will focus specifically on the phenomenologically important Anderson model in one dimension and utilize modern supersymmetry-based techniques to give a compact analytic treatment of its properties geared toward practical applications. This will be followed by an economical and effective numerical implementation for solving the resulting integral equations.\vspace{1mm}

{\it The auxiliary supervector model and its solution.}---
Consider a slight generalization of the standard Anderson model: take a real symmetric tridiagonal $L\times L$ Hamiltonian matrix $H$ whose diagonal entries $H_{ii}$ (the on-site potentials) are independent random variables governed by the probability distribution $h(x)$, and the subdiagonal entries $H_{i+1,i}=H_{i,i+1}$ (the hopping amplitudes) are independent random variables governed by the probability distribution $g(x)$. The rest of the entries are zero. For future use, we introduce the following Fourier transforms of the probability distributions of the nonzero entries:
\beq\label{Fourier}
\tilde h(k)\equiv \int dx\, h(x)\,e^{-ikx},\quad \tilde g(k)\equiv \int dx\, g(x)\,e^{-ikx}.
\eeq

To compute the DoS, which is the density of eigenvalues $E_n$ of $H$, we use the standard combination of the resolvent representation and Gaussian supervector integrals, as reviewed in \cite{laplace}. Namely, we write \cite{SP}
\allowdisplaybreaks
\begin{align}
&\rho(E)\equiv\frac1L\sum_{n=1}^L\langle\de(E-E_n)\rangle=\frac{-1}{\pi L}\Im\sum_{n=1}^L\langle[E-H+i0]^{-1}_{nn}\rangle\nn\\
&=\!\frac1{\pi L}\!\Im\sum_{n=1}^L\!\int\![dX] (i\bar x_n x_n)\,e^{iE\sum_kX^\dagger_k X_k} \langle e^{-i\sum_{kl} H_{kl} X^\dagger_k X_l}\rangle,\nn\\
&\hspace{1cm}[dX]\equiv\prod_{k=1}^L dX_k,\qquad dX\equiv \frac1\pi d\bar x  dx  d\bar\xi d\xi.\label{reslv}
\end{align}
\nodisplaybreaks
where each $X_k$ with $k=1..L$ is a two-component complex supervector $X_k\equiv (x_k, \xi_k)$, with commuting (bosonic) $x_k$ and anticommuting (fermionic) $\xi_k$, meaning that
$\xi_k\xi_l=-\xi_l\xi_k$, $\xi_k\bar\xi_l=-\bar\xi_l\xi_k$, $\bar\xi_k\bar\xi_l=-\bar\xi_l\bar\xi_k$,
satisfying the standard Berezin integration rules
\beq
\forall k:\,\int d\xi_k=\int d\bar\xi_k=0,\,\,\, \int d\xi_k \xi_k=\int d\bar\xi_k\bar\xi_k=1, 
\eeq
while the scalar product $X_1^\dagger X_2$ is defined as
\beq\label{XdX}
X_1^\dagger X_2\equiv \bar x_1x_2+\bar \xi_1\xi_2.
\eeq
The general idea behind (\ref{reslv}) is that, with (\ref{XdX}) substituted, it factorizes into two Gaussian integrals: over $x$ and over $\xi$. The first one gives $[E-H]^{-1}_{nn}/\det[E-H]$, while the second gives $\det[E-H]$, leaving behind altogether the desired expression \cite{gaussint}. The purpose of introducing the anticommuting variables is 
precisely to make these determinant factors cancel.

Since (\ref{reslv}) is factorized over the entries of $H$, we can straightforwardly average it over the $H$-ensemble defined by the probabilities $h(x)$ and $g(x)$,  yielding
\begin{align}
\rho(E)=&\frac1{\pi L}\Im\sum_{n=1}^L\int[dX] \,(i\bar x_n x_n)\prod_{l=1}^L\tilde h(X^\dagger_l X_l)\,e^{iEX^\dagger_l X_l} \nn\\
 &\hspace{1.5cm}\times\prod_{k=1}^{L-1} \tilde g(X^\dagger_k X_{k+1}+X^\dagger_{k+1}X_k).\label{superv}
\end{align}
The next step in the standard treatments in the literature \cite{Efetov, nucleus,FMsigma,Mirlin} is to introduce a supermatrix made from bilinears of $X$ an $X^\dagger$ and to rewrite (\ref{superv})
as a nonlinear $\sigma$-model for this supermatrix. This is typically followed by taking a continuum limit of the discrete chain and looking for the condensation loci of the supermatrix. It is not obvious how to recover the DoS of the original model from the endpoint of these derivation, where a continuum limit has already been taken. We will follow a different, and more straightforward, route.

The integral (\ref{superv}) is easily recognized as a sort of discrete-time quantum mechanics of a `superparticle' described by the configuration variable $X$. 
To make this manifest, we introduce the following shorthand
\beq
F(k,E)\equiv \sqrt{\tilde h(k) \,e^{iEk }},
\eeq
and the `transfer' operator $\mathcal{T}$ acting on (wave)functions of supervectors:
\begin{align}\label{Tdef}
\mathcal{T}\Psi(X)\equiv F(X^\dagger X,E)& \int dX' \,\tilde g(X^\dagger X'+X'^\dagger X)\\
&\hspace{1cm}\times F(X'^\dagger X', E) \Psi(X').\nn
\end{align}
Then, (\ref{superv}) is identically rewritten as \cite{transfer}
\begin{align}\label{supervT}
\rho(E)=\frac1{\pi L}\Im\sum_{n=1}^L\int dX \,&F(X^\dagger X,E)\,\mathcal{T}^{n-1}\mathcal (i\bar x x)\\
&\hspace{1cm}\times\mathcal{T}^{L-n}F(X^\dagger X,E).\nn
\end{align}
If $L$ is large, in most terms of this sum, both $n-1$ and $L-n$ are large numbers, and hence $\mathcal{T}^{n-1}$ and $\mathcal{T}^{L-n}$ will simply project on the eigenfunction of $\mathcal{T}$ with the largest eigenvalue (in terms of the magnitude of its modulus), and multiply by powers of the said eigenvalue, which we denote $\lambda_*$, while the corresponding normalized eigenfunction is $\Psi_*(X)$. In other words, we assume that $\mathcal{T}^m$ can be approximated by a projector at large $m$:
\beq
\mathcal{T}^m f(X)\underset{\sssty m\to\infty}{\sim} \lambda_*^m \Psi_*(X)\int dX' \Psi_*(X') f(X').
\eeq
(Note that $\mathcal{T}$ is symmetric and in general non-Hermitian.)
 This results in the following estimate at large $L$:
\begin{align}\label{rhoinsertion}
\rho(E)=&\frac{\lambda_*^{L-1}}{\pi}\Im \left(\int dX \Psi_*(X)\,F(X^\dagger X,E) \right)^2  \\
&\hspace{3cm}\times\int dX (i\bar x x)\Psi^2_*(X).\nn
\end{align}
The $(\cdots)^2$ factor comes from the endpoints of the chain, while the last factor comes from the insertion $(i\bar xx)$.
In order for $\rho(E)$ to have a well-defined $L\to\infty$ limit, one must have $\lambda_*=1$. A way to ensure it is to assume that $\Psi_*$ is supersymmetric, in other words, $\Psi_*(X)=\Psi_*(X^\dagger X)$. This assumption immediately simplifies many integrals due to the identity \cite{textbooks, sup2}
\beq\label{Fdef}
\int dX f(X^\dagger X)= f(0)
\eeq
valid \cite{oneline} for any supersymmetric function $f(X^\dagger X)$ that decays at infinity.
Note, in particular, that this formula shows that $\int dX \Psi_*(X^\dagger X)\mathcal{T}\Psi_*(X^\dagger X)/\int dX \Psi_*^2=1$ which means that our assumption $\lambda_*=1$ follows from the supersymmetry of $\Psi_*$. Furthermore, the normalization $\int dX \Psi^2_*=1$ translates into 
\beq\label{Psinorm}
\Psi_*(0)=1.
\eeq
We also have $\int dX \Psi_*(X^\dagger X)\,F(X^\dagger X,E) =1$, since $\tilde h(0)=1$ by the probability normalization of $h$. With all of this,
\beq\label{rhoX}
\rho(E)= \frac1\pi \Im\int dX\, (i\bar x x)\,\Psi^2_*(X^\dagger X),
\eeq
where $\Psi_*$ satisfies
\begin{align}\label{PsiXX}
\Psi_*(X^\dagger X)=&F(X^\dagger X,E) \int dX' \,\tilde g(X^\dagger X'+X'^\dagger X)\\ 
&\hspace{2cm}\times F(X'^\dagger X',E) \, \Psi_*(X'^\dagger X')\nn
\end{align}
and the normalization (\ref{Psinorm}). 

A side remark is that, especially in a system with localized eigenstates, one expects global quantities like the DoS to be insensitive to the boundary conditions when the system is large. Most eigenstates are localized somewhere in the bulk and know nothing about the boundary conditions imposed at the endpoints of the chain. Representations like (\ref{rhoinsertion}) make this feature mathematically manifest, since different boundary conditions would have changed the irrelevant supersymmetric integral appearing in the first line \cite{PBC}, but not the $\bar x x$ insertion appearing in the second line that defines the final answer (\ref{rhoX}).

Since $\Psi_*$ is a function of one composite variable $X^\dagger X$, all the integrals can be simplified. To this end, we introduce
$r\equiv \bar x x$, $r'\equiv\bar x' x'$, $\alpha\equiv\arg{x'}-\arg{x}$, so that $dX'=dr'd\alpha d\bar\xi'd\xi'/{2\pi}$,
and rewrite (\ref{PsiXX}) as
\begin{align}\label{PsiXX}
&\Psi_*(r+\bar\xi\xi)=F(r+\bar\xi\xi,E)\int\frac{dr'd\alpha}{2\pi} d\bar\xi' d\xi'\,\\
&\times\tilde g(2\sqrt{rr'}\cos\alpha +\bar\xi\xi'+\bar\xi'\xi)\,  F(r'+\bar\xi'\xi',E)\,\Psi_*(r'+\bar\xi'\xi').\nn
\end{align}
By anticommutativity of $\bar\xi$ and $\xi$, $\Psi_*(r+\bar\xi\xi)=\Psi_*(r)+\bar\xi\xi\del_r\Psi_*(r)$, since $\xi^2=\bar\xi^2=0$. As everything is symmetric under superrotations, it suffices to match the pieces independent of $\xi$ and $\xi'$ on the two sides of (\ref{PsiXX}), and then the remaining pieces will match automatically (this can also be verified explicitly \cite{FM}). Extracting such $(\xi,\bar\xi)$-independent pieces yields
\begin{align}\label{Psistar}
\Psi_*(r)=&-F(r,E) \int_0^\infty \hspace{-2mm}dr' \,  \del_{r'}\left[F(r',E)\Psi_*(r')\right]\\
&\hspace{2cm}\times\frac1{2\pi}\int_0^{2\pi}\hspace{-2mm}d\alpha  \,\tilde g(2\sqrt{rr'}\cos\alpha),\nn
\end{align}
where the minus sign arises from the need to reorder the anticommuting $\xi$'s before taking the Berezin integral.
Using the standard integral formula for the Bessel function $J_0(x)=(2\pi)^{-1}\int_0^{2\pi} d\alpha\, e^{ix\cos\alpha}$ and (\ref{Fourier}), we can rework the last line into 
\begin{align}
\Psi_*(r)=&-\sqrt{\tilde h(r)}\,e^{iEr/2} \int_0^\infty \hspace{-2mm}dr' \,\del_{r'}\!\left[\sqrt{\tilde h(r')}\,e^{iEr'/2}\Psi_*(r')\right] \nn\\
&\hspace{2cm}\times\int dx\, g(x) J_0(2x\sqrt{rr'}).\label{PsiJ}
\end{align}
Similar manipulations \cite{J1} convert (\ref{rhoX}) into
\beq\label{rhostar}
\rho(E)=\frac1\pi\Re\int_0^\infty \hspace{-2mm}dr \,\Psi^2_*(r,E).
\eeq

Overall, most generally, one must solve (\ref{PsiJ}) for $\Psi_*(r,E)$ numerically at each value of $E$, as described below, keeping in mind the normalization $\Psi_*(0)=1$, and then reconstruct $\rho(E)$ from (\ref{rhostar}).

{\it Fully solvable cases.}--- We can check how the general formulas (\ref{PsiJ}-\ref{rhostar}) operate for the simple cases where explicit analytic solutions were previously known. First, there is the trivial case without disorder, where $ h=\de(x)$, $g=\de(x-1)$, and hence
\beq
\Psi_*(r)=-e^{iEr/2 }\int_0^\infty \hspace{-2mm}dr' \,  \del_{r'}\!\left[\Psi_*(r') e^{iEr'/2 }\right]J_0(2\sqrt{rr'}).
\eeq
To solve this, we undo the step leading to the Bessel function by writing 
$J_0(x)=(2\pi)^{-1}\int_0^{2\pi} d\alpha\, e^{ix\cos\alpha},$
and then introduce a 2d plane $(y_1,y_2)$ via $y_1\equiv\sqrt{r}\cos\alpha$, $ y_2\equiv\sqrt{r}\sin\alpha$, $dr d\alpha/2=dy_1 dy_2\equiv dy_{12}$,
so that
$$
\Psi_*(r)\!=\!-\frac{e^{iEr/2 }}{\pi}\!\!\int \!dy_{12}\,e^{2i\sqrt{r}y_1} \del_{r'}\!\!\left[\Psi_*(r') e^{iEr'/2 }\right]_{r'={y_1^2+y_2^2}}.
$$
The right-hand side evidently converts exponentials of something linear in $r$ to exponentials of something linear in $r$. Namely, assume
\beq\label{Psiexp}
\Psi_*(r)=e^{-\alpha r}.
\eeq
Then, $e^{-\alpha r}=e^{iEr/2-r/(\alpha-iE/2)}$, $\alpha^2+{E^2}/4-1=0$, and hence
$\alpha(E)={\sqrt{4-E^2}}/2$.
 From (\ref{rhostar}) and (\ref{Psiexp}),
\beq
\rho(E)=\Re\frac{1}{2\pi\alpha(E)}=\frac{1}{\pi\sqrt{4-E^2}}.
\eeq
This is the correct $L\to\infty$ limit of the DoS of the uniform nearest-neighbor hopping chain, with its Van Hove singularities at the edges.

The derivation goes through in essentially the same way for the solvable Lloyd model \cite{lloyd} with the Cauchy distribution of the on-site potentials: $h(x)=[\pi W(1+x^2/W^2)]^{-1}$ and $g(x)=\de(x-1)$, so that $\tilde h(k)=e^{-W|k|}$. Then, we get
\begin{align}\label{PsiLloyd}
\Psi_*(r)=&-\frac{e^{(iE-W)r/2 }}{\pi}\int dy_1 dy_2\, e^{2i\sqrt{r}y_1} \\
&\hspace{1cm}\times\del_{r'}\!\left[\Psi_*(r')e^{(iE-W)r'/2 }\right]_{r'={y_1^2+y_2^2}}.\nn
\end{align}
With (\ref{Psiexp}), this gives
$\alpha^2-(iE-W)^2/4-1=0$, $\alpha(E)={\sqrt{4+(W-iE)^2}}/2$ and
\beq
\rho(E)=\Re\frac{1}{\pi\sqrt{4+W^2-E^2-2iW E}},
\eeq
which is the correct expression for the Lloyd model DoS.

{\it Numerical implementation.}---
One cannot solve (\ref{PsiJ}) in terms of elementary functions in general, but effective numerical approaches are available. It is natural to approximate $\Psi_*$ as a truncated expansion in some complete basis, and we will use a power series expansion multiplied by a suitably chosen weight. The treatment here is guided by the numerical solutions of related (and more complicated) nonlinear integral equations in \cite{hammer,mixreg}. The following identity \cite{Sz} is useful for decoupling the dependences on $r$ and $r'$ in (\ref{PsiJ}) and minimizing the number of integrals one has to compute numerically:
\beq\label{Besselid}
J_0(2\sqrt{vw})=e^{-w}\sum_{n=0}^\infty \frac{w^n\,L_n(v)}{n!},
\eeq
where $L_n(v)$ are Laguerre polynomials orthogonal on $[0,+\infty)$ with respect to the measure $dv\, e^{-v}$. We use this identity in (\ref{PsiJ}) while setting $v=x^2r'/C$, and $w=Cr$, where $C$ at this point is an arbitrary adjustable parameter that we can use to control the numerical performance. 
This gives
\begin{align}\label{PsiJexpand}
&\Psi_*(r)=-F(r,E)\,e^{-Cr}\sum_{n=0}^\infty \frac{(Cr)^n}{n!} \\
&\times\int_0^\infty \hspace{-2mm}dr' \,  \del_{r'}\!\left[F(r',E)\Psi_*(r')\right] \int dx\, g(x) L_n(x^2r'/C).\nn
\end{align}
Using the explicit formula for Laguerre polynomials,
$L_n(x)=\sum_{m=0}^n{n \choose m}{(-x)^m}/{m!}$
we then write
\begin{align}
\Psi_*(r)=&-F(r,E)\,e^{-Cr}\sum_{n=0}^\infty \frac{(Cr)^n}{n!} \sum_{m=0}^n{n \choose m}\frac{g_{m}}{(-C)^mm!}\nn\\
&\hspace{1cm}\times\int_0^\infty \hspace{-2mm}dr' \,  r'^m \,\del_{r'}\!\left[F(r',E)\Psi_*(r')\right].\label{Psimomnt}
\end{align}
where we have introduced the even moments of the distribution $g(x)$ given by
\beq\label{gmmnt}
g_{m}\equiv  \int dx\, g(x) \,x^{2m}.
\eeq
We now expand $\Psi_*$ according to
\beq\label{Psidecomp}
\Psi_*(r)=F(r,E)\,e^{-Cr}\sum_{m=0}^M \beta_m r^m,
\eeq
with a suitable truncation at order $M$. We similarly truncate the sum on the right-hand side of (\ref{Psimomnt}) at $n=M$. Equating the coefficients of the individual terms $r^m$, we get
\begin{align}
\beta_n=& -\frac{C^n}{n!} \sum_{m=0}^n{n \choose m}\frac{g_{2m}}{(-C)^mm!}\\
&\hspace{1cm}\times\int_0^\infty \hspace{-2mm}dr\,  r^m\, \del_{r}\!\left[\sum_{k=0}^M \beta_k r^k\tilde h(r)\,e^{iEr}e^{-Cr}\right].\nn
\end{align}
We can separate out the term with $m=0$ and apply integration by parts in the remaining terms to get
\beq\label{betaeq}
\beta_n=\frac{C^n}{n!}\left[\beta_0+\sum_{k=0}^M \beta_k \sum_{m=1}^n{n \choose m}\frac{g_{m}\tilde h_{m+k-1}}{(-C)^m(m-1)!}\right],
\eeq
where 
\beq\label{hmmnt}
\tilde h_m\equiv\int_0^\infty \hspace{-2mm}dr\,  r^{m}\,\tilde h(r)\,e^{iEr }e^{-Cr}.
\eeq
The normalization (\ref{Psinorm}) translates into
\beq\label{betanorm}
\beta_0=1.
\eeq
One can then assume this value of $\beta_0$ and solve (\ref{betaeq}) for the remaining $\beta$'s. Once all the $\beta$'s are known, $\rho(E)$ is recovered from (\ref{rhostar}) as
\beq\label{rhobeta}
\rho(E)=\frac1\pi\Re\!\!\sum_{m,k=0}^M \!\!\beta_m\beta_k\int_0^\infty \hspace{-3mm}dr\,r^{m+k} \,\tilde h(r)\,e^{iEr }e^{-2Cr}.
\eeq

For given $g$ and $\tilde h$, one must first compute (\ref{gmmnt}) and (\ref{hmmnt}), and then solve (\ref{betaeq}) subject to (\ref{betanorm}), followed by reconstruction of $\rho(E)$ from (\ref{rhobeta}). 

To validate these results, we can explore a few concrete examples: First, without disorder in the hoppings, $g(x)=\de(x-1)$, consider $h(x)$ taken as a uniform distribution between $-W$ and $W$, so that
\beq\label{uniform}
\tilde h(k)=\frac{\sin Wk}{Wk},\qquad g(x)=\de(x-1).
\eeq
Second, still without any disorder in the hoppings, $g(x)=\de(x-1)$, take $h(x)$ to be a Gaussian distribution of width $W$ so that 
\beq
h(x)=\frac1{W\sqrt{2\pi}} e^{-x^2/2W^2},\qquad\tilde h(k)=e^{-k^2W^2/2}.\label{Gaussian}
\eeq
Third, one can try an asymmetric combination of Gaussian distributions for the on-site potentials keeping $g(x)=\de(x-1)$:
\begin{align}
&h(x)=\frac1{3W\sqrt{2\pi}}\left[ e^{-(x-1)^2/2W^2}+2e^{-(x+1/2)^2/2W^2}\right]\nn \\ 
&\tilde h(k)=\frac{e^{-k^2W^2/2}}3\left(e^{-ik}+2e^{ik/2}\right).\label{asymm}
\end{align}
Finally, take the case with Gaussian variables on all three diagonals:
\beq\label{gaussall}
\begin{split}
&h(x)=\frac1{\sqrt{2\pi}} e^{-x^2/2},\qquad\tilde h(k)=e^{-k^2/2},\\
&g(x)=\frac1{\sqrt{2\pi}} e^{-x^2/2},\qquad g_m=(2m-1)!!.
\end{split}
\eeq

Implementing the process outlined above in Python with NumPy and SciPy using $C=1$, $M=20$ produces results in excellent agreement with numerically sampled eigenvalue histograms, as seen in Fig.~\ref{numerics}. A basic script is provided in the supplemental material \cite{sppl}.
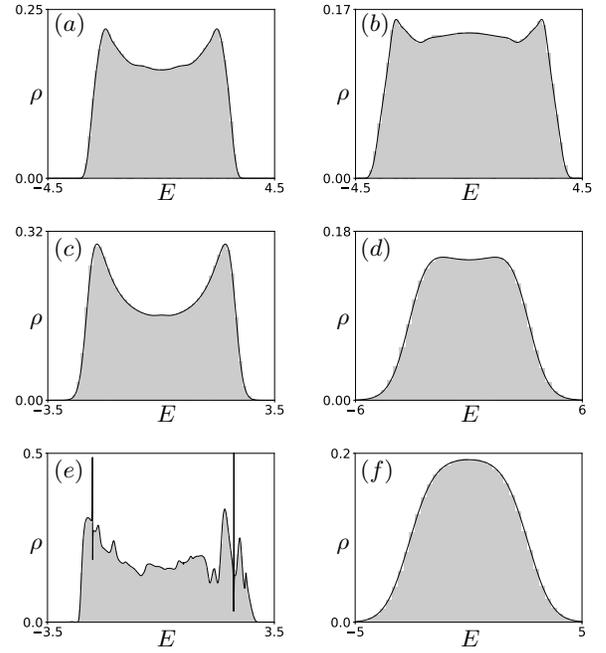
\begin{figure}[t]
\includegraphics[width = 0.45\linewidth]{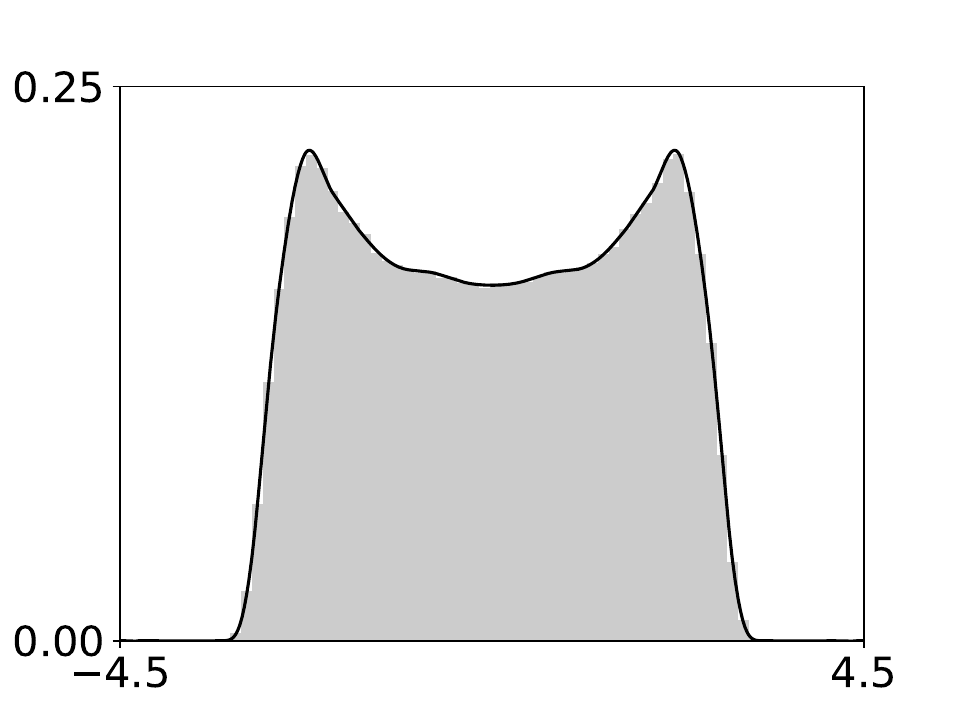}\hspace{2mm}\includegraphics[width = 0.45\linewidth]{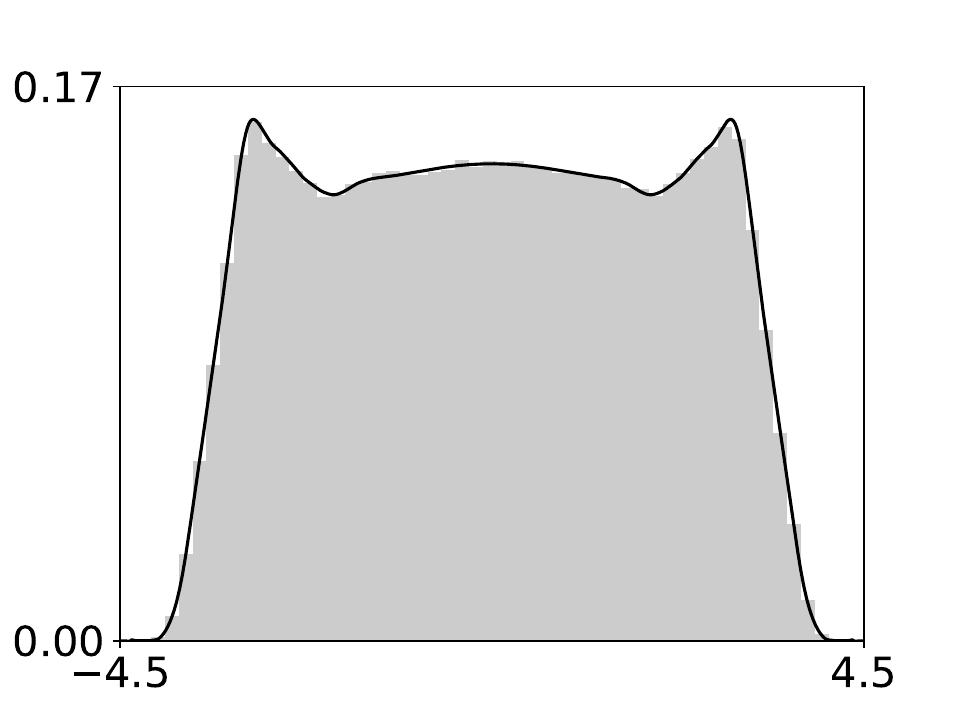}\\
\includegraphics[width = 0.45\linewidth]{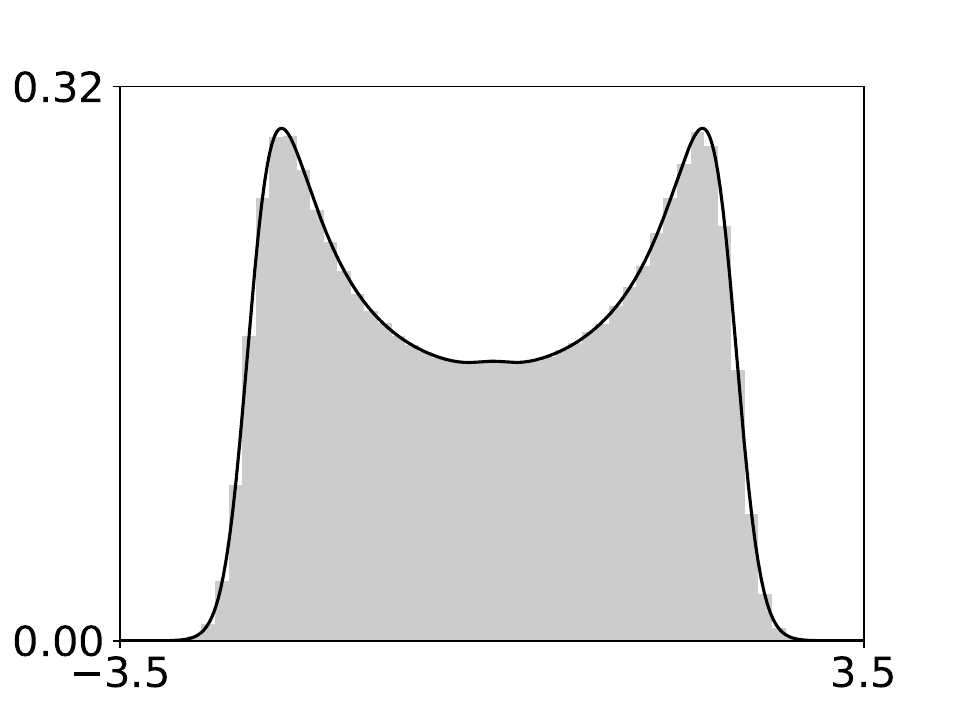}\hspace{2mm}\includegraphics[width = 0.45\linewidth]{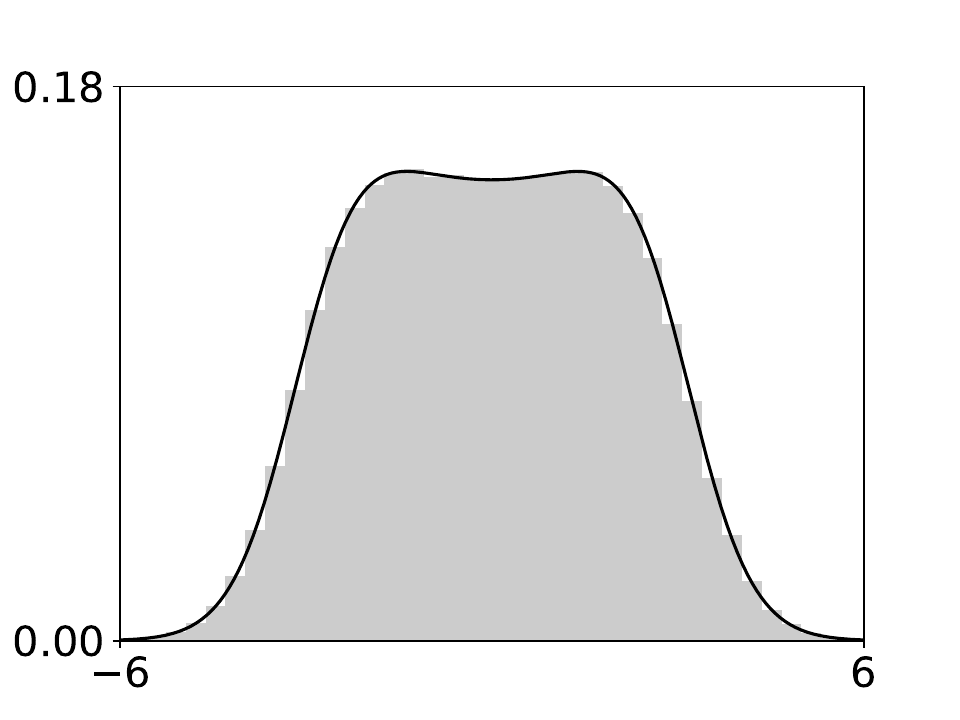}\\
\hspace{1mm}\includegraphics[width = 0.45\linewidth]{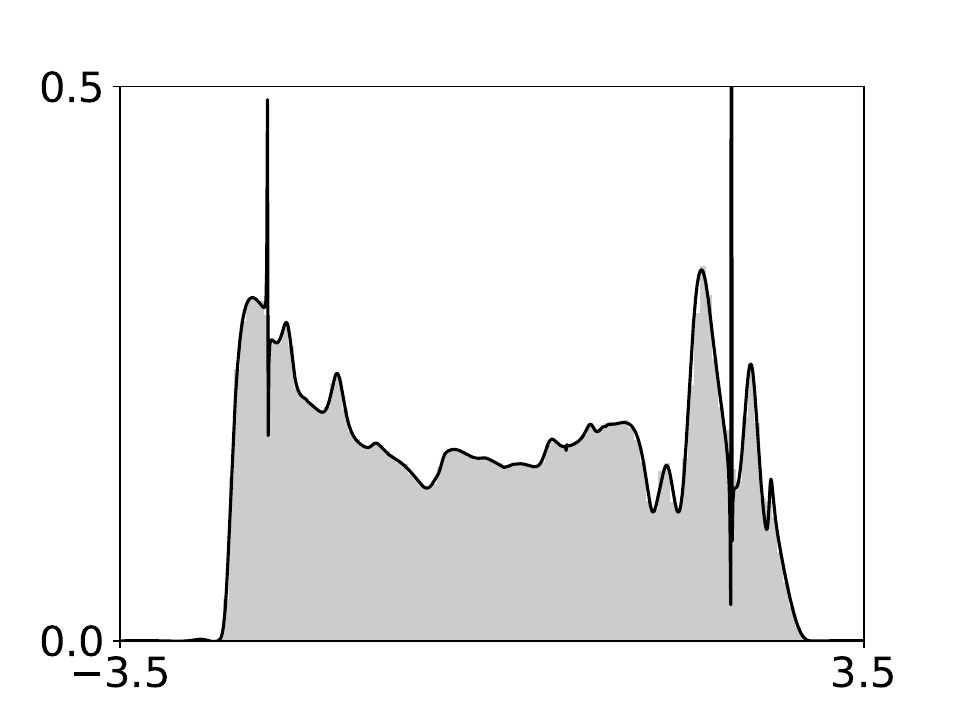}\hspace{2mm}\includegraphics[width = 0.45\linewidth]{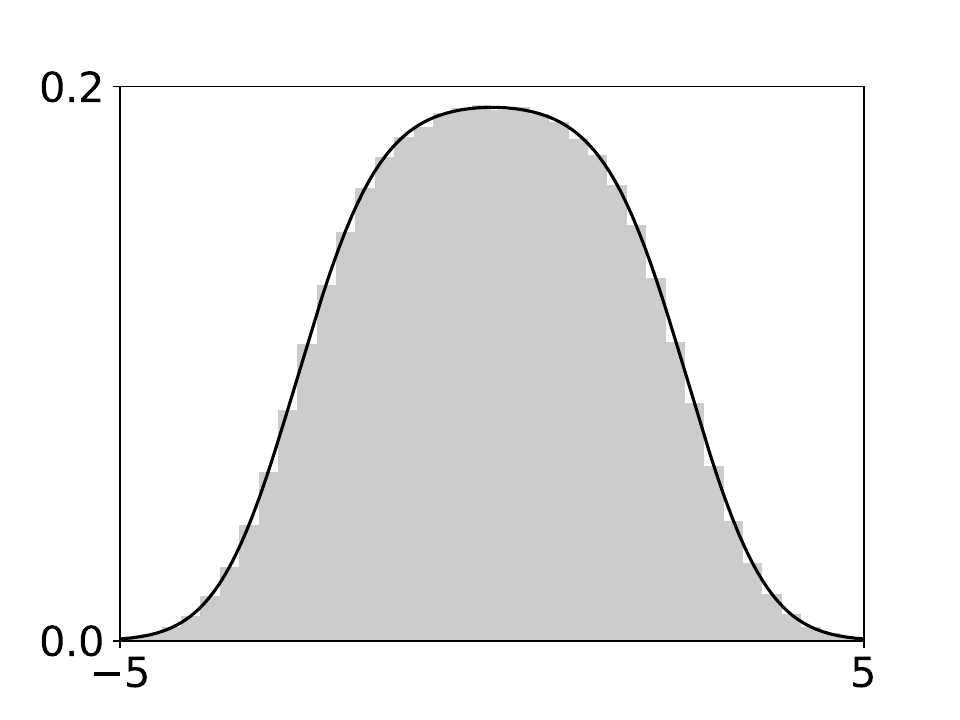}
\begin{picture}(0,0)
\put(-223,207){$\rho$}
\put(-175,169){$E$}
\put(-214,233){$(a)$}
\put(-107,207){$\rho$}
\put(-60,169){$E$}
\put(-98,233){$(b)$}
\put(-223,122){$\rho$}
\put(-175,85){$E$}
\put(-214,148){$(c)$}
\put(-107,122){$\rho$}
\put(-60,85){$E$}
\put(-98,148){$(d)$}
\put(-223,37){$\rho$}
\put(-175,0){$E$}
\put(-214,64){$(e)$}
\put(-107,37){$\rho$}
\put(-60,0){$E$}
\put(-98,64){$(f)$}
\end{picture}
\caption{Analytic predictions obtained by solving (\ref{betaeq}) and evaluating (\ref{rhobeta}), represented as black curves, superposed on top of the histograms generated by sampling 100 random matrices at $L=10000$ for each model specification: {\bf (a)} for (\ref{uniform}) with $W=1.5$; {\bf (b)} the `cat' distribution for (\ref{uniform}) with $W=2.5$; {\bf (c)} for (\ref{Gaussian}) with $W=0.5$; {\bf (d)} for (\ref{Gaussian}) with $W=1.5$; {\bf (e)} for (\ref{asymm}) with $W=0.1$: the extravagant shape of this curve, accurately captured by the analytic framework presented here, is understood from approaching a discrete distribution of on-site potentials at small values of $W$---the DoS is known to develop singularities for discrete distributions \cite{discr}; {\bf (f)} for (\ref{gaussall}).}
\label{numerics}
\end{figure}

{\it Localization length.}--- The DoS of the Anderson model is an interesting quantity in its own right, and it finds real-life applications, for instance, in the physics of disordered polymer chains \cite{DoSpoly}. This said, the most frequently discussed aspect of the Anderson model is the localization properties of its eigenstates at energy $E$, rather than in the density of their corresponding eigenvalues. Using special properties of tridiagonal matrices, Thouless has given a formula, however, that expresses the localization exponent of eigenvectors at energy $E$ through the DoS \cite{thouless}, which can be written in the language used here as
\beq
\lambda(E)=\int dE'\,\rho(E')\log|E-E'|-\int dx\, g(x)\log|x|.
\eeq
This formula then immediately converts the analytic expression (\ref{rhostar}) derived here to an explicit result for the localization exponent \cite{locint, IPR}.\vspace{1.5mm}

To sum up, application of supersymmetry-based methods for disorder averaging leads to an explicit analytic representation for the density-of-states of the 1d Anderson model with arbitrary probability distributions for the on-site potentials and nearest-neighbor hopping amplitudes: one has to solve, at each value of the eigenstate energy, the 1d linear integral equation (\ref{PsiJ}) and recover the density-of-states from (\ref{rhostar}). A key step in deriving these results is recognizing that multiple applications of the supersymmetric transfer operator (\ref{Tdef}) effectively act as a projector on a special supersymmetric eigenfunction corresponding to eigenvalue 1. The equation defining this eigenfunction is then reduced to (\ref{PsiJ}). The final output is in the form of elementary integral equations, and does not depend in any way on the supersymmetric formalism that had been used at the intermediate steps. These equations are amenable to easy numerical solution and the results show perfect agreement with  numerical sampling of large matrices.

Having analytic control over a model in general simplifies the process of handling it in practice, and one can foresee a number of applications.
For instance, one can use the representations developed here for searches of disorder distributions with interesting properties, or for reverse-engineering disorder
distributions with the desired density-of-states. The techniques developed here should transfer to the block-Anderson model \cite{Molinari}, which is a subject for future work. Recent years have seen prominent emergence of random tridiagonal matrices in situations distinct from their original applications to 1d transport of particles and waves. Thus, acting with the Lanczos algorithm on quantum Hamiltonians and Liouvillians leads to tridiagonal representations of general quantum dynamics in the context of Krylov complexity program \cite{K,spread,Krev}, generating renewed interest in the analysis of random tridiagonal matrices and their localization properties \cite{Kint,trid,nocorr}.
On a very different front, studies of statistical thermodynamics of classical Lax-integrable systems in \cite{Spohn,Doyon} have given rise to models of random tridiagonal matrices linked with the thermodynamic Bethe ansatz. I hope that the techniques presented here will provide a valuable tool for these diverse pursuits.\vspace{4mm}

\begin{acknowledgments}
I am indebted to Pawat Akara-pipattana, Thip Chotibut, Ben Craps, Weerawit Horinouchi, Dima Kovrizhin, Gabriele Pascuzzi and Napat Poovuttikul for discussions and collaboration on related subjects.
The analytic developments presented here have been triggered by a conversation with Luca Molinari, which is cordially acknowledged, as well as his subsequent guidance through aspects of the past literature and comments on the manuscript.
This work has been supported by by the  C2F program at Chulalongkorn University and by NSRF via grant number B41G680029.
\end{acknowledgments}\vspace{2mm}

\onecolumngrid

\subsection*{Integral equations for one-dimensional disordered problems}

\twocolumngrid

The workflow presented in the main text starts with applying a Gaussian supervector representation for the resolvent of a disordered 1d Hamiltonian, leading in the long chain limit to a linear integral equation that controls the density-of-states (DoS). A compact numerical implementation is then provided for solving this equation, showing an excellent agreement with stochastic averages over ensembles of random tridiagonal matrices.

\hspace{3mm}Bibliographic studies reveal a few strands in the past literature where integral equations are discussed in relation to the Anderson model and its relatives. Closest methodologically to the treatment in the main text is the work of Campanino and Klein \cite{transfer} whose follow-up is mostly in the mathematical literature and focuses on formal analytic properties of the DoS rather than on practical recipes for its evaluation. Additionally, there is a line of work originating from Dyson's classic paper \cite{Dyson} on disordered oscillator chains. Curiously, this paper predates most of Dyson's extensive work on random matrices as well as Anderson's original paper on localization. In this paper, the method of moments is used, which is mathematically rigorous, but the derivations are very convoluted in comparison with the compact supersymmetry-based derivations presented here. The output is in the form of linear integral equations, though the algebraic presentation is rather different from here. In this work and its follow-up in \cite{Schmidt, Halperin, Forrester}, the specification of disorder within the tridiagonal matrices differs from the independent entries typical of random hopping models with onsite potentials (for instance, for oscillator chains, one randomizes the masses and spring constants which, in turn, determine the entries of tridiagonal matrices whose DoS one wants to construct).

\hspace{3mm}A very different branch of work starts with classic papers of Abou-Chacra, Anderson and Thouless on Bethe lattices with random onsite potentials \cite{ACAT}. Some follow-up can be found in \cite{nuclphysb,DerridaRodgers,BST}. Bethe lattices are infinite regular trees where each vertex is connected to $d$ other vertices (and there are no loops). If one picks a vertex and starts moving away from it, at each step, the path will split into $d-1$ branches. When $d\ge 3$, this geometry has infinite-dimensional features since the number of sites at distance less than a given value grows exponentially, and hence faster than any power. The special case $d=2$, on the other hand, corresponds to a 1d chain. Bethe lattices are evoked for studying infinite-dimensional mean-field physics, and one in general can reduce their description to nonlinear integral equations with polynomial nonlinearities of degree $d-1$. The derivations, however, only use the treelike character of the graph and hold in the special case $d=2$ as well. In this case, the integral equations controlling the model become linear and connect to the considerations in this paper. While the presentation in \cite{ACAT} is not recognizably similar to the treatment in the main text, it was later recast in terms of supersymmetric integrals in \cite{nuclphysb} in a manner close to the presentation here. In \cite{DerridaRodgers}, an alternative derivation is given using more elementary mathematics (there is some disagreement about signs between this paper and \cite{nuclphysb}).

\hspace{3mm}Overall, the main message of the present treatment is that the supersymmetric transfer operator techniques that have historically appeared in \cite{transfer}, and were mostly used for addressing formal analyticity questions, can be converted into an economical tool for computations within the Anderson model in a way that is accessible and must be made available for a range of applications.

\onecolumngrid\vspace{5mm}

\subsection*{Numerical implementation}

\indent The Python script below reproduces Fig.~1c. For a fast run, moderate values are chosen for the number of $E$-points for evaluation of the analytic curve $\rho(E)$ and the dimension and number of samples for the random matrices. These can be increased for higher-quality output. The script can be adapted straightforwardly for other cases treated in the main text.\vspace{2mm}

\begin{verbatim}
import numpy as np
import scipy as sp
import matplotlib.pyplot as plt

W=0.5
C=1
M=21
L=1000
sample=100

E=np.linspace(-3.5,3.5,100)
rho=np.zeros((len(E)))

fac=np.zeros((M))
fac[0]=1.0
fac[1]=1.0
for n in range(2,M):
 fac[n]=fac[n-1]*n

for i in range(len(E)):
 #precomputing the integrals in (31) and (33)
 hh=[sp.integrate.quad(lambda x: x**n*np.exp(1j*E[i]*x-C*x-(W*x)**2/2), 0, 20.0/C, limit=500,\
     complex_func=True)[0] for n in range(2*M)]
 hhh=[sp.integrate.quad(lambda x: x**n*np.exp(1j*E[i]*x-2*C*x-(W*x)**2/2), 0, 20.0/C, limit=500,\
     complex_func=True)[0] for n in range(2*M)]
 #precomputing the coefficients of the linear system (30) with beta0 set to 1
 A=np.zeros((M-1,M-1),dtype='complex_') #indices 0,1... correspond to beta1, beta2,...
 b=np.zeros((M-1),dtype='complex_')
 for n in range(M-1):
  A[n,n]=-1
  b[n]=-C**(n+1)*(1+np.sum([fac[n+1]*hh[m-1]/((-C)**m*fac[m]*fac[n+1-m]*fac[m-1])\
                             for m in range(1,n+2)]))/fac[n+1]
  for k in range(M-1):
   A[n,k]+=C**(n+1)*np.sum([fac[n+1]*hh[m+k]/((-C)**m*fac[m]*fac[n+1-m]*fac[m-1])\
                             for m in range(1,n+2)])/fac[n+1]
 #solving(30)
 beta=[1.0]+np.linalg.solve(A,b).tolist()
 #applying (33)
 rho[i]=np.real(np.sum([beta[k]*np.sum([beta[m]*hhh[m+k] for m in range(M)])\
                        for k in range(M)]))/np.pi
 
#sampling random matrices
eiglist=[]
for rept in range(sample):
 H=np.zeros((L,L))
 for i in range(L):
  H[i,i]=W*np.random.normal()
  if i<L-1:
   H[i,i+1]=1
   H[i+1,i]=1
 eiglist+=np.linalg.eigvalsh(H).tolist()

plt.hist(eiglist,bins=50,density=True,color='0.8')
plt.plot(E,rho,'k')
plt.xlim(-3.5,3.5)
plt.ylim(0,0.32)
plt.show()
\end{verbatim}


\begin{thebibliography}{99}
\bibitem{gaussint}The precise formulas used here are as follows: for any Hermitian matrix $M$, $\int\left[\prod\frac{d\bar x_n dx_n}\pi\right]\,\bar x_p x_q\,e^{i\sum_{kl}M_{kl}\bar x_k x_l}=[-iM]^{-1}_{pq}/\det[-iM]$, and $\int\left[\prod d\bar \xi_n d\xi_n\right]\,e^{i\sum_{kl}M_{kl}\bar \xi_k \xi_l}=\det[-iM]$. This is then applied to $M=E-H$.
\bibitem{oneline}Here is a one-line proof: in the polar coordinates $x=\sqrt{r}e^{i\alpha}$, one can write $\int dX\, f(X^\dagger X)=(2\pi)^{-1}\int drd\alpha d\bar \xi d\xi\, f(r+\bar\xi\xi)=
\int  dr d\bar \xi d\xi (f+\bar\xi\xi\del_rf)=-\int_0^\infty dr\,\del_r f(r)=f(0)$ if $f$ decays at infinity.
\bibitem{PBC}If we had periodic boundary conditions, for example, an extra $\tilde g$-factor would have appeared in (\ref{superv}), coupling the $L$th site to the first and resulting in $\mathcal{T}^L$ in (\ref{supervT}) instead of two separate insertions of $\mathcal{T}$ for the open boundary conditions. Once supersymmetry has been taken intro account, the resulting expression (\ref{rhoX}) will remain unchanged.
\bibitem{J1}We could integrate by parts in (\ref{PsiJ}), which would have replaced the kernel based on $J_0(2x\sqrt{rr'})$ with a kernel based on $J_1(2x\sqrt{rr'})/\sqrt{rr'}$. Such kernels are familiar from the theory of sparse matrices with randomly placed entries \cite{BR,FM}.
The difference, however, is that our integral equation (\ref{PsiJ}) is linear, while the equations in \cite{BR,FM} include exponential nonlinearities. Our situation is, therefore, simpler. (Additionally, the integral equations of \cite{BR,FM} have been later rederived using mathematically rigorous methods in \cite{Khetal,reslarge}. This makes it likely that the integral equations developed here can be rederived in a mathematically rigorous way as well.)
\bibitem{IPR}More refined quantities describing localization have been considered in the literature, including variants of the inverse participation ratio (IPR), see~\cite{Mirlin}. These quantities can be expressed as averages of products of matrix inverses, which generalizes the structure of (\ref{reslv}), and should be tractable using the transfer operator techniques.
\end{thebibliography}
\end{document}